# In-situ creation and control of Josephson junctions with a laser beam

W. Magrini[1,2,4], S. V. Mironov[3], A. Rochet[1,2], Ph. Tamarat[1,2], A. I. Buzdin[4] and B. Lounis[1,2]

[1]*University Bordeaux, LP2N, F-33405 Talence, France*
[2]*Institut d'Optique & CNRS, LP2N, F-33405 Talence, France*
[3]*Institute for Physics of Microstructures RAS, Nizhny Novgorod, GSP-105, Russia*
[4]*University Bordeaux, LOMA, F-33405 Talence, France*

brahim.lounis@u-bordeaux.fr

**We propose the use of a laser beam tightly focused on a superconducting strip to create a Josephson junction by photothermal effect. The critical current of this junction can be easily controlled by the laser intensity. We show that a periodic modulation of the intensity substantially changes the dynamic properties of the junction and results in the appearance of Shapiro steps without microwave radiation. The experimental realization of optically driven Josephson junctions may open a way for the ultra-fast creation and switching of complex patterns of superconducting devices with tunable geometry and current-phase relations.**

A Josephson junction (JJ), consisting of two superconductors separated by a non-superconducting material (the so-called weak link) or a solid superconductor with an artificial constriction, is one of the key elements of modern cryogenic electronics,[1] quantum computing systems, [2,3] ultrasensitive electric and magnetic sensors[4] and other types of quantum devices. The non-dissipative electric current $j_s$ flowing through the junction is coupled with the phase difference $\varphi$ between the gap potentials inside the superconducting electrodes, and the corresponding current-phase relation (CPR) $j_s(\varphi)$ determines all main static and dynamic characteristics of a Josephson system.[5,6] For basic JJs with an insulating or a normal metal weak link, the CPR is primary controlled by temperature and/or external magnetic field. Cooling the sample switches the CPR from a sinusoidal function to a linear one, while the magnetic field damps the critical current (the maximal current which can flow through the junction without dissipation) and induces peculiar Fraunhofer oscillations.[5] Even richer physics arise when the superconducting electrodes are separated by a ferromagnetic layer. [7] In contrast to the usual JJ with the zero phase $\varphi$ in the ground state, ferromagnets allow to create $\pi$-junctions with a spontaneous ground state phase $\varphi = \pi$ [8,9] or even $\varphi_0$-junctions with $\varphi = \varphi_0 \neq 0, \pi$. The latter possibility requires a strong spin-orbit coupling[10,11] or a large spin splitting of the electron energy bands due to the exchange field.[12,13] By applying an external magnetic field and changing the exchange field orientation, one can tune $\varphi_0$, which opens new perspectives for the elements of the rapid single flux-quantum logics.[1,14,15] However, in all existing types of tunable JJs, the weak link needs to be embedded into the system during the fabrication process and can hardly be tuned afterwards. The versatility and performance of Josephson devices would greatly benefit from simple methods to tune *in situ* the CPR of the built-in JJs and eliminate fabrication inhomogeneities.

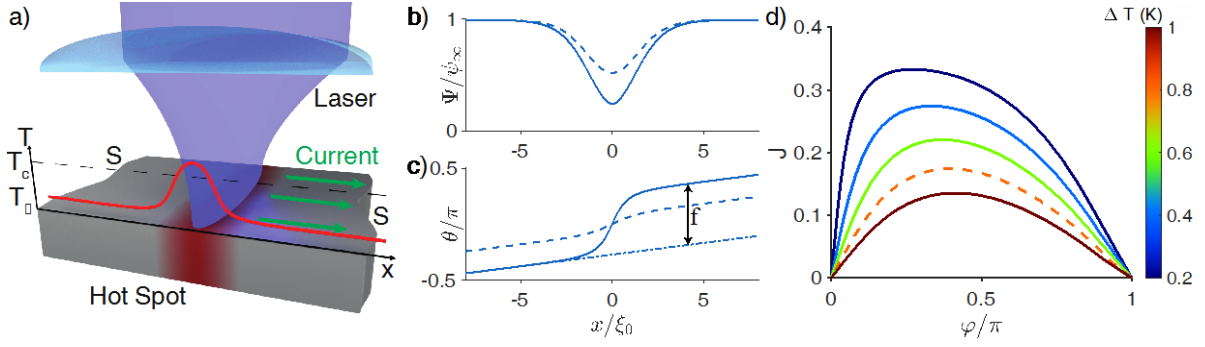

Figure 1: a) The concept of a laser induced Josephson junction. A laser beam focused on a superconducting strip creates a hot spot which damps the order parameter and induces a weak link and hence a JJ. The red curve illustrates the temperature profile along the strip. b) Profiles of the order parameter along the laser-induced JJ. The solid (dashed) line corresponds to the stable (unstable) state. Here the dimensionless current J is set to 0.1 and the coherence length of the superconductor $\xi_0$ to 100 nm (relevant for an Al strip). c) Profiles of the superconducting phase θ along the strip, demonstrating the appearance of the Josephson phase difference φ. d) Current-phase relations of the JJ for different laser intensities resulting in different temperature elevations ΔT. For the calculations we took the base temperature $T_\infty = 0.4$ K and the critical temperature $T_c = 1.2$ K (relevant to Al).

We have recently shown that a tightly focused laser beam induces a strong thermal gradient in the superconductor which can be used to manipulate single flux quanta [16]. The laser locally heats the superconductor and creates a sub-micron hot spot with a temperature rise which can be controlled by the beam power. Here we propose the concept of a JJ created *in situ* at any desired region of the superconductor and tailored by photothermal effect, using a short-wavelength laser beam tightly focused on a superconductor strip [see Fig. 1(a)].

In conventional JJs, the size of the weak link should be of the order of the zero-temperature superconducting coherence length $\xi_0$, which ensures a substantial critical current. Because of the diffraction limit, the minimal size of the hot spot induced by the laser cannot be smaller than several hundreds of nm, which exceeds $\xi_0$ for most superconductors. Here we show that in the presence of a non-uniform temperature profile with a maximal temperature slightly above the superconducting transition temperature $T_c$, the effective coherence length which controls the Josephson current becomes strongly increased at the interface between the normal metal (N) region induced at the hot-spot and its peripheral superconducting (S) area. Therefore, the laser-induced local temperature rise can be tuned to generate a weak link as narrow as the effective coherence length and, hence, a Josephson junction. The fast (sub-ns) thermal response time of the superconductor enables an ultra-fast creation of the JJ and a real-time control of the Josephson transport by light at GHz frequencies.

To uncover the physics beyond this concept we calculate the CPR of the laser-induced JJ within the Ginzburg-Landau (GL) approach.[17] Our calculation procedure is similar to the one suggested in Ref. 18, where the CPR of the JJ formed by the non-uniform concentration of impurities affecting $T_c$ was analyzed. For simplicity we restrict ourselves to the case of a one-dimensional superconductor, assuming that the thickness of the slab is much smaller than the hot spot radius $R$ and the superconducting coherence length $\xi$. The free energy $F$ of the superconducting condensate and the local current density $j_s$ are determined by the profile of the order parameter $\Psi(x)$ as

$$F = \int_{-\infty}^{\infty} \left\{ \alpha[T(x) - T_c]|\Psi|^2 + \gamma|\partial_x \Psi|^2 + \frac{\beta}{2}|\Psi|^4 \right\} dx \quad (1),$$

$$j_s = -\frac{2\pi i \gamma}{\Phi_0}(\Psi^* \partial_x \Psi - \Psi \partial_x \Psi^*) \quad (2),$$

where $x$ is the coordinate along the superconductor with $x = 0$ in the center of the laser beam, $T(x)$ is the profile of temperature inside the hot spot, $T_c$ is the critical temperature of the superconducting transition, $\alpha$, $\beta$ and $\gamma$ are positive GL coefficients. For simplicity we assume the temperature profile $T(x)$ to have a Gaussian shape $T(x) = T_\infty + \Delta T \exp(-x^2/2R^2)$ with a base temperature $T_\infty < T_c$ and a temperature elevation $\Delta T$ set by the laser beam intensity. Indeed for a thin superconducting film the temperature can be considered as uniform in the yz-planes, insuring a uniform current density $j_s$ in the whole superconducting stripe. In order to find the profile $\Psi(x)$ realizing the minimum of Eq.(1), it is convenient to represent the order parameter in the form $\Psi(x) = f(x)\psi_\infty e^{i\theta(x)}$, where $\psi_\infty^2 = \alpha(T_c - T_\infty)/\beta$ is the absolute value of the order parameter at $x \to \pm\infty$ in the absence of current, $f(x)$ is a real function, and $\theta$ is the superconducting phase. Introducing the dimensionless current $J = \frac{\Phi_0 j_s \beta}{4\pi\sqrt{\alpha^3\gamma(T_c-T_\infty)^3}}$ and the dimensionless position $\tilde{x} = x\sqrt{\frac{\alpha}{\gamma}(T_c - T_\infty)}$ from Eq.(2), one finds that $\partial_{\tilde{x}}\theta = J/f^2(\tilde{x})$. Then the Ginzburg-Landau equation for the function $f(\tilde{x})$ obtained from the variation of Eq.(1) with respect to $\Psi^*$ gives:

$$-\partial_{\tilde{x}}^2 f + \tau(\tilde{x})f + f^3 + \frac{J^2}{f^3} = 0 \qquad (3),$$

where $\tau(\tilde{x}) = [T(\tilde{x}) - T_c]/(T_c - T_\infty)$, and $f(\tilde{x})$ satisfies the boundary conditions $\partial_{\tilde{x}}f(\tilde{x} \to \infty) = 0$. This equation for $f(\tilde{x})$ admits two distinct solutions, a stable one and an unstable one for each value of $J$, as long as that the applied current is lower than the critical current $J_c$ of the junction. To find them for a given value of $J$, we numerically determine the values of the function $f(0)$ at $\tilde{x} = 0$ for which the solution of Eq.(3) satisfies the boundary condition at $\tilde{x} \to \infty$. The stable solution corresponds to the so-called zero-state of the JJ, which is characterized by a zero Josephson phase difference in the ground state (for $J = 0$) while the unstable one corresponds to the energetically unfavorable $\pi$-state. The Eq.3 allows us to calculate the order parameter profile along the junction, which is plotted in Fig. 1(b). The phase profile along the junction is determined from the expression $\partial_{\tilde{x}}\theta = J/f(\tilde{x})^2$. In the absence of laser radiation, the profile $\theta(\tilde{x})$ is linear provided that the current does not exceed the depairing current value. However, the weak link formed in the hot-spot region with the damped order parameter favors a large additional variation of the superconducting phase, which is typical for JJs[5]. The resulting profile $\theta(\tilde{x})$ is displayed in Fig. 1(c) and presents a crossover between two linear asymptotes. The shift between them is in fact the Josephson phase difference $\varphi$ which can be expressed through the $f(\tilde{x})$ profile:

$$\varphi = 2J \int_0^\infty [f^{-2}(\tilde{x}) - f_\infty^{-2}]d\tilde{x} \qquad (4).$$

Here we take into account the symmetry relation $f(-\tilde{x}) = f(\tilde{x})$ and introduce the value $f_\infty = f(\tilde{x} \to \infty)$ which depends on the current $J$. By solving these equations for various applied currents, we are able to calculate the current-phase relation $J(\varphi)$ of the optically induced JJ for a given laser power. Fig. 1(d) shows the current-phase relations of the JJ for different amplitudes of the laser-induced temperature elevation $\Delta T$. The increase in laser power naturally results in the damping of the JJ critical current, which provides an effective tool for the real-time optical control of the superconducting transport through the junction.

In the situation where the temperature of the hot-spot center is above $T_c$, the effective coherence length which controls the Josephson transport is determined by the thermal gradient in the vicinity of the contour with $T = T_c$, which separates the normal region in the hot-spot center from the superconducting periphery. At small distances $\delta$ from this S/N boundary, the local temperature $T(\delta)$ can be approximated by $T(\delta) \simeq T_c + \nabla T \cdot \delta$ where $\nabla T$ is the temperature gradient at the S/N boundary. Then the effective coherence length is self-consistently defined as $\xi_{eff} \simeq \xi_0/\sqrt{\frac{\nabla T}{T_c}\delta}$ at the distance $\delta = \xi_{eff}$ from the S/N boundary. This leads to $\xi_{eff} \simeq \sqrt[3]{L_T \xi_0^2}$ where $L_T = T_c/\nabla T \gg \xi_0$ is the typical length of temperature variation. Such local increase of the coherence length enables the use of a vast range of superconducting materials, including those with relatively small $\xi_0$, for the realization of the laser induced JJ.

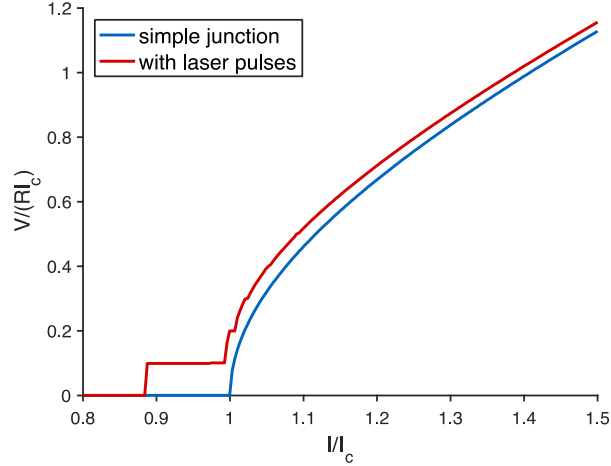

Figure 2: Comparison of the current-voltage characteristics of a classical JJ (blue curve) and a superconducting strip irradiated by gaussian laser pulses (red curve). The curves are calculated from the resistively shunted junction model, using $I_{c,1}/I_{c,0} = 0.2$, $\tau = t_0/10$ and $\Omega = 2\pi/t_0 = 0.1\omega_0$ with $\omega_0 = 4\pi e R_N I_{c,0}/\hbar$, where $e$ is the electron charge and $R_N$ is the normal resistance of the device.

Due to their dynamic properties, Josephson junctions are also known to show the emergence of current steps (called Shapiro steps) in their current-voltage characteristic under the influence of the microwave radiation. Remarkably, a periodic modulation of the laser intensity substantially changes the dynamic properties of the JJ and results in the appearance of Shapiro steps, even without microwave radiation. To show this, we consider a JJ with a sinusoidal CPR under the effect of a train of laser pulses which periodically damp the order parameter and thus the JJ critical current. It is worth noting that the typical response time of the superconducting condensate lies in the ps range, which ensures that the supercurrent will adiabatically follow the temperature variations at the ns time-scale. Each pulse is assumed to have a Gaussian temporal profile, so that the critical current depends on time as

$$I_c(t) = I_{c,0} - I_{c,1} \sum_{n=1}^{\infty} e^{-\frac{(t-nt_0)^2}{2\tau^2}},$$

where $I_{c,1}$ is the amplitude of the critical current variation, $t_0$ is the period of the pulse train, and $\tau$ is the pulse duration. Under application of a d.c. voltage $V_0$ to the junction, the Josephson phase becomes time dependent with $\varphi(t) = \varphi_0 + \omega_f t$, where $\omega_f = 2eV_0/\hbar$ and $\varphi_0$ is the initial phase. Expanding the expression for $I_c(t)$ into the Fourier series, we obtain for the superconducting current:

$$I(t) = \left\{ I_{c,0} - \frac{\Omega\tau}{2\sqrt{\pi}} I_{c,1} \left[ 1 + 2 \sum_{n=1}^{\infty} e^{-\left(\frac{n\Omega\tau}{2}\right)^2} \cos(n\Omega t) \right] \right\} \sin[\varphi_0 + \omega_f t], \quad (5)$$

where $\Omega = 2\pi/t_0$. This current contains d.c. components provided that $\omega_f = \pm n\Omega$, which corresponds to the emergence of a series of Shapiro spikes with amplitudes $\delta I_n = \frac{\Omega\tau}{\sqrt{\pi}} I_{c,1} e^{-\left(\frac{n\Omega\tau}{2}\right)^2}$. The laser radiation thus enables an active control of the dynamic properties of the junction. Fig. 2 compares the I-V curves of such a device with and without periodic laser radiation in the frames of the resistively shunted junction model[5] and shows the formation of steps on the I-V curve induced by the laser pulse train. In contrast to the conventional Shapiro steps which start around the zero current, these steps appear close to the critical current.

Moreover, the amplitude and spacing of these steps can be finely tuned by changing the repetition rate, pulse width and/or laser power.

An additional microwave radiation of the frequency $\omega_r$ applied to the JJ naturally adds an a.c. voltage component $V_r = V_1 \cos(\omega_r t)$ between the superconducting electrodes. The equation of the non-stationary Josephson effect[5] determines the time evolution of the phase difference $\varphi(t) = \varphi_0 + \omega_f t + a \sin(\omega_r t)$, where $a = (\omega_f/\omega_r)(V_1/V_0)$. In the presence of microwave radiation, the conventional Shapiro steps appear for $\omega_f = m\omega_r$ ($m$ is integer). The corresponding d.c. components of the current are given by $I = I_{c0} J_m(mV_1/V_0) \sin\varphi_0$, where $J_m$ is a Bessel function of the $m$-th order. To sum up, in the presence of microwave radiation, the I-V characteristics of the laser driven JJ have two sets of Shapiro steps at d.c. voltages $V_0$ corresponding to the frequencies

$$\omega_f = m\omega_r \pm n\Omega. \quad (6)$$

To propose a realistic experimental configuration for the realization of the optically driven Josephson junction, one may consider a superconducting strip made from a type-I superconductor such as Al (with $T_c = 1.2\ K$ and a large coherence length $\xi_0 \sim 100\ nm$) deposited on top of the sapphire substrate. This substrate has a thermal conductivity much larger than that of the superconductor, allowing the minimization of the hot spot size. The thickness of the strip should be chosen small enough to ensure a uniform in-depth heating by the laser beam and large enough to enable measurable Josephson critical currents. Solving the heat equation numerically for a 5 $nm$-thick Al layer with temperature $T_\infty = 0.4\ K$ on top of 1 $mm$-thick sapphire substrate, we find that a laser beam with power ~ 1 mW and wavelength 400 nm focused onto a spot with radius $\sim 100\ nm$ produces a local increase of temperature up to 1 K in the superconductor. As shown in Fig 1.(d), such a temperature elevation is high enough for in situ control of the JJ critical current on a large scale, from zero to almost the depairing current of the superconducting strip.

**Conclusion**

In this work, we theoretically showed that a focused laser beam irradiating a superconducting strip creates a Josephson junction by a local temperature increase and the subsequent damping of the superconducting gap. The critical current of such an optically induced Josephson junction is controlled by the beam power and thus easily tunable up to GHz frequencies. We also predict that periodic modulation of the critical current with a train of laser pulses results in the onset of Shapiro steps in the current-voltage characteristic of the junction, even without microwave radiation. The proposed concept of laser induced JJ may open the way for tunable Josephson systems with unconventional radiation properties. Moreover, this work can be extended to the in-situ generation and coherent control of complex patterns of *identical* Josephson junctions. A similar approach was used to create Josephson patterns in Bose-Einstein condensates[19]. Using structured optical illumination, one may design and produce new macroscopic quantum states by optically imprinting a phase pattern onto a raw superconducting film. This may provide a substantial advance in the fabrication and tailoring of large Josepshon arrays consisting of $10^5$-$10^6$ JJs that are widely used, e.g., in d.c. and a.c. voltage standards [20].

**Acknowledgments**

The authors thank A. Mel'nikov and H. Majedi for useful discussions. This work was supported by the French National Agency for Research "ANR SUPERTRONICS" and "ANR OPTOFLUXONICS", the EU COST CA16218 Nanocohybri, the Région Nouvelle Aquitaine, Idex Bordeaux (LAPHIA Program) and the Institut universitaire de France. S. M. acknowledges financial support from the Russian Science Foundation under Grant No. 18-72-10027, the Foundation for the advancement of theoretical physics "BASIS", the Russian Presidential Scholarship SP-3938.2018.5 and the Russian Foundation for Basic Research under Grant No. 18-02-00390.